\documentclass[pra,aps,twocolumn,showpacs,nofootinbib,superscriptaddress]{revtex4}

\usepackage{graphicx}
\usepackage{subfigure}  
\usepackage{amssymb}
\usepackage{amsmath}
\usepackage{bm}

\def\jcp#1#2#3{J.~Chem.~Phys.~{\bf #1},\ #2\ (#3)}

\def\pra#1#2#3{Phys.~Rev.~A~{\bf #1},\ #2\ (#3)}

\def\prl#1#2#3{Phys.~Rev.~Lett.~{\bf #1},\ #2\ (#3)}

\def\bea{\begin{eqnarray}}
\def\eea{\end{eqnarray}}
\def\be{\begin{equation}}
\def\ee{\end{equation}}
\def\etal{{\it et al.}}

\newcommand{\ket}[1]{|#1\rangle}                                     

\def\ddpar{\partial}

\begin{document}

\title{CrRb: a molecule with large magnetic and electric dipole moments}
\author{Z. Pavlovi\'{c}}
\email[]{pavlovic@phys.uconn.edu}
\affiliation{\mbox{ITAMP, Harvard-Smithsonian Center for Astrophysics, 
60 Garden Street, Cambridge, Massachusetts 02138, USA}}
\affiliation{Department of Physics, University of Connecticut, Storrs, 
Connecticut  06269-3046, USA}
\author{H. R. Sadeghpour}
\email[]{hrs@cfa.harvard.edu}
\affiliation{\mbox{ITAMP, Harvard-Smithsonian Center for Astrophysics, 
60 Garden Street, Cambridge, Massachusetts 02138, USA}}
\author{R. C\^{o}t\'{e}}
\email[]{rcote@phys.uconn.edu}
\affiliation{Department of Physics, University of Connecticut, Storrs, 
Connecticut  06269-3046, USA}
\author{B. O. Roos}
\email[]{deceased.}
\affiliation{Department of Theoretical Chemistry, University of Lund, S-221 00 Lund, Sweden}
\date{\today}

\begin{abstract}
We report calculations of Born-Oppenheimer potential energy curves 
of the chromium-rubidium heteronuclear molecule ($^{52}$Cr$^{87}$Rb), and  
the long-range dispersion coefficient for the interaction between  ground state  Cr and Rb atoms. 
Our calculated van der Waals coefficient ($C_6=1770$ a.u.) has an expected error of 3\%. 
The ground state $^6\Sigma^+$ molecule at its equilibrium separation has a permanent electric 
dipole moment of $d_e (R_e=3.34 \text{ \AA}) = 2.90\ D$. We investigate the hyperfine and 
dipolar collisions between trapped Cr and Rb atoms, finding elastic to 
inelastic cross section ratio of $10^2{-}10^3$.
\end{abstract}
\maketitle
\section{Introduction}
The creation and manipulation of ultracold  polar molecules in their ground state is a 
growing endeavor in atomic and  molecular  physics.
Rubidium is the workhorse of ultracold atomic physics. It was the first atom to be Bose condensed  
and it remains a favorite atom for simulations of correlated and many-body physics, and for 
forming cold molecules. The chromium atom has also been Bose condensed and it is the only 
atomic species for which strong dipolar interaction has been observed \cite{pfauBEC}. 
Dipole moments, by virtue of their long-range $1/R^3$ interaction, where $R$ is
 the separation between atoms, can be tuned by external fields with increasing
  precision to control atomic interactions 
with applications to precision measurements, molecular physics and information processing.
\par
CrRb is a heteronuclear molecule, and may therefore possess a sizable permanent electric dipole 
moment. In its ground electronic state, it has magnetic moment of five Bohr magnetons
 ($5\mu_B$), so it can be 
magnetically tuned. In its most abundant form (84\%), $^{52}$Cr has no nuclear spin, $i=0$, but its 
fermionic isotope $^{53}$Cr (9.5\% abundance) has nuclear spin, $i = 3/2$, that couples to
 electronic spin, $s=3$, to produce a number of hyperfine levels. Therefore, bosonic-fermionic mixtures 
of CrRb can be formed with large electric and magnetic dipole moments, with potentially 
interesting applications for degenerate dipolar Fermi gases and spinor physics. 
\par
There is no spectroscopic information available for the CrRb molecule. A two-species magneto-optical trap (MOT) for Cr 
and Rb was realized in 2004 \cite{pfauCrRb}, in which some $4\times 10^6$ $^{52}$Cr atoms and 
$3\times 10^6$ $^{87}$Rb atoms were loaded. Cr is known to have large 
inelastic two-body spin-flip losses, so it cannot be maintained in a MOT 
\cite{pfauCr01,pav05}. Owing to its large magnetic moment, magnetic trapping (MT) of Cr is 
possible \cite{pfauCr01}, which allows the atoms to be trapped in its lowest high-field seeking 
state with total electronic spin $s=3$ and  its projection $m_s=-3$.  In the Rb-MOT + Cr-MT configuration, Hensler 
\etal \cite{pfauCrRb} measured the two-body loss rate constant to be 
$\beta_{\rm RbCr} \sim 1.4 \times 10^{-11}$ cm$^{-3}$/s and 
$\beta_{\rm CrRb} \sim  10^{-10}$ cm$^{-3}$/s, where the former refers to the loss due to 
the introduction of Cr into the Rb-loaded MOT, and the latter refers to loading of the Cr-MT first. 
The latter loss coefficient is about an order of magnitude larger than the former loss 
coefficient, because in the more shallow Cr-MT, interspecies dipolar interactions lead 
more quickly to spin depolarizing collisions, hence depletion of the trap.
\par
In this {\it ab initio} work, we analyze the electrostatic interaction of Rb and Cr atoms in their 
ground states, obtain Born-Oppenheimer (BO) potential ground state energy curves, the long-range 
dispersion $C_6$ coefficient, and present a value for the electric dipole moment of the CrRb 
molecule. We use the resulting potential energy curves to calculate the elastic 
and relaxation cross sections, hyperfine and dipolar Fano-Feshbach resonances 
which can be used to tune the interactions. 

\section{Numerical Calculations and Collisional Results}

The calculations of the BO potential energy curves and the permanent electric dipole moments 
were performed with the multiconfiguration complete active space self-consistent field 
(CASSCF/CASPT2) method  \cite{roos87,roos92}. 
The basis set was of 
VQZP quality with the primitives obtained from the relativistic ANO-RCC basis 
set ($7s6p4d3f2g1h$ for Cr and $8s7p4d2f1g$ for Rb)\cite{roos04,roos05}. Scalar relativistic effects 
are included in the calculations using the Douglas-Kroll-Hess Hamiltonian, as is 
standard in the MOLCAS software. 
Two active spaces were used. The first comprised the Cr $3d$ and $4s$ orbitals and the Rb $5s$, 
thus seven orbitals with seven active electrons. In the second set of calculations, a second 
set of $3d$ orbitals was added to describe the $3d$ double shell effect: 12 orbitals 
with seven electrons. All calculations were performed with the MOLCAS-7 quantum chemistry 
software \cite{molcas}. 
\par
There are two electronic states which correlate to Rb ($6s \ ^2S$) and 
Cr($3d^54s \ ^7S$) ground states,  $^6\Sigma^+$ and $^8\Sigma^+$. 
At large separation, $R$, these curves are well described by $-C_6/R^6$, where $C_6$ is the 
van der Waals (vdW) coefficient.
We calculate  $C_6$ using the Casimir-Polder 
\cite{msd94} integral
\begin{align}
 C_6 = \frac{3}{\pi} \int_{0}^{\infty} 
  \alpha_{\rm Cr}(i \omega) \alpha_{\rm Rb} (i\omega) d\omega,
\end{align}
\begin{figure}[t]
\includegraphics[width=0.99\columnwidth]{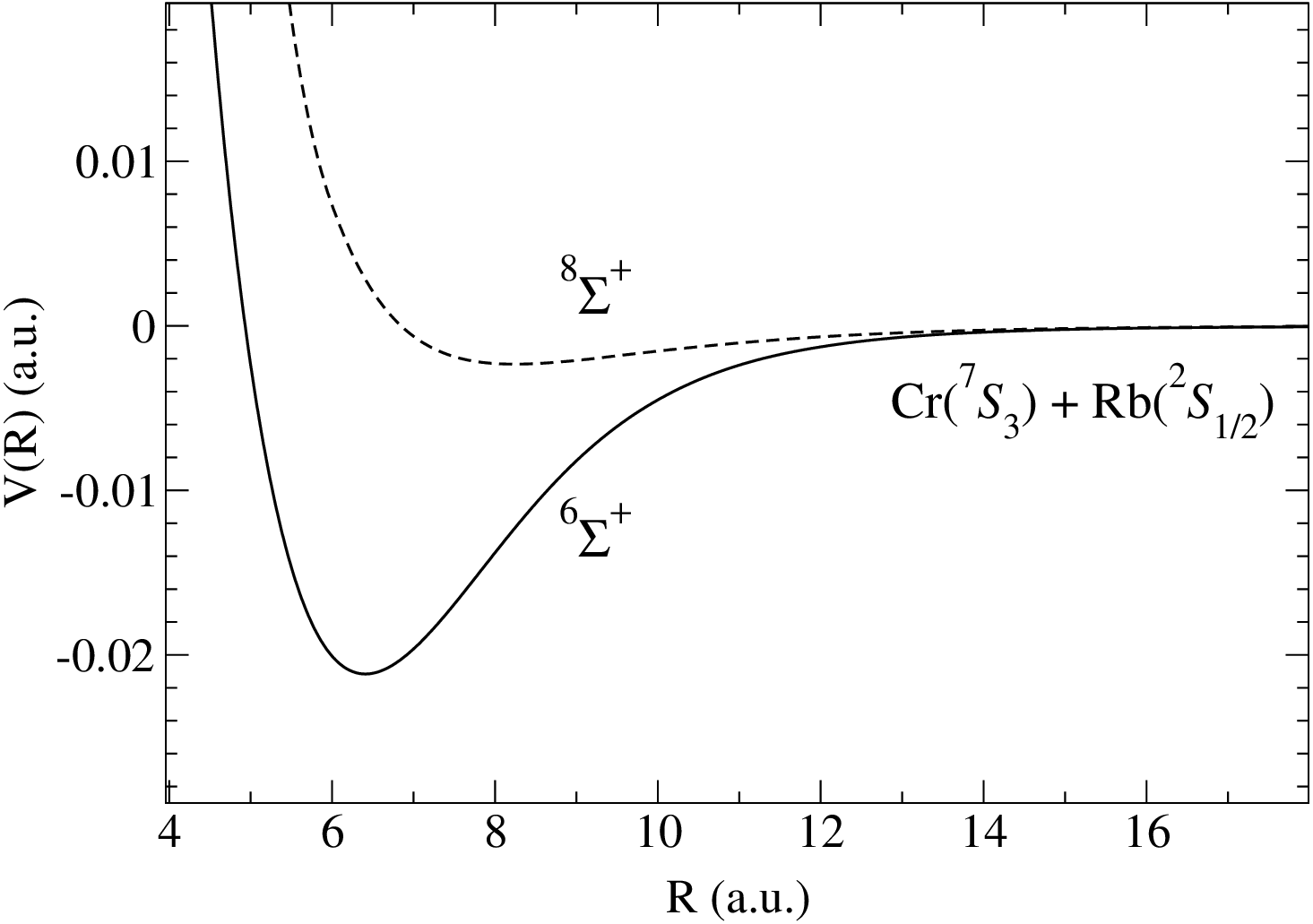}
\caption{The $^6\Sigma^+$ and $^8\Sigma^+$ potential energy curves for CrRb, 
correlating to Cr($^7S_3$) and Rb($^2S_{1/2}$) ground states.}
\label{pot}
\end{figure}
where  the dynamic polarizability of each atom, $\alpha(i \omega)$, is given by
\begin{align}
 \alpha(i \omega) = \sum_{\lambda} 
 \frac{f_{0\lambda}}{\epsilon_{\lambda}^2 + \omega^2}.
\end{align}
The calculation of  the dynamic polarizability requires knowledge of the oscillator 
strengths $f_{0\lambda}$ between the atomic ground state of energy $\epsilon_0$ 
and atomic excited states of energy $\epsilon_{\lambda}$.
The summation is understood to also include  integration over continuum. Our  calculated value of 
$C_6$ for the Cr $+$ Rb system, $1770$ a.u., which we 
believe to be accurate within 3\%, is obtained from the highly-accurate values of 
the Rb dynamic polarizability at imaginary frequencies \cite{der99} and the 
recently accurate values for the Cr dipole polarizability \cite{pav04,pfau05}.
This should be compared to the values  of $C_6({\rm Rb}_2) = 4691$ a.u. 
\cite{der99,msd94}, and 
$C_6 ({\rm Cr}_2) = 770$ a.u., \cite{pav05}.
Both $^6\Sigma^+$ and $^8\Sigma^+$ curves share the same $C_6$ 
coefficient.
\par
In Fig.~\ref{pot}, we plot the BO potential energy curves for CrRb in the ground 
electronic states. The equilibrium separation for the molecule in the ground state 
$^6\Sigma^+$  is $R_e=6.31$ a.u.
 The value for the dipole moment at the equilibrium distance for 
the $^6\Sigma^+$ ground state molecule is $d_e = 2.90$ D ($1.14$ a.u.), which is reasonably 
large. The calculated static dipole polarizability for the molecule is 
$\alpha_{\rm CrRb} (0) = 701$ a.u. We should mention that 
$\alpha_{\rm Rb} (0) = 319$ a.u. \cite{bederson} 
 and $\alpha_{\rm Cr}(0)= 86$ a.u. \cite{pav05}. 
The molecular value is 40\% larger than the sum of atomic polarizabilities.
\par
\begin{figure} [t]
\includegraphics[width=0.99\columnwidth]{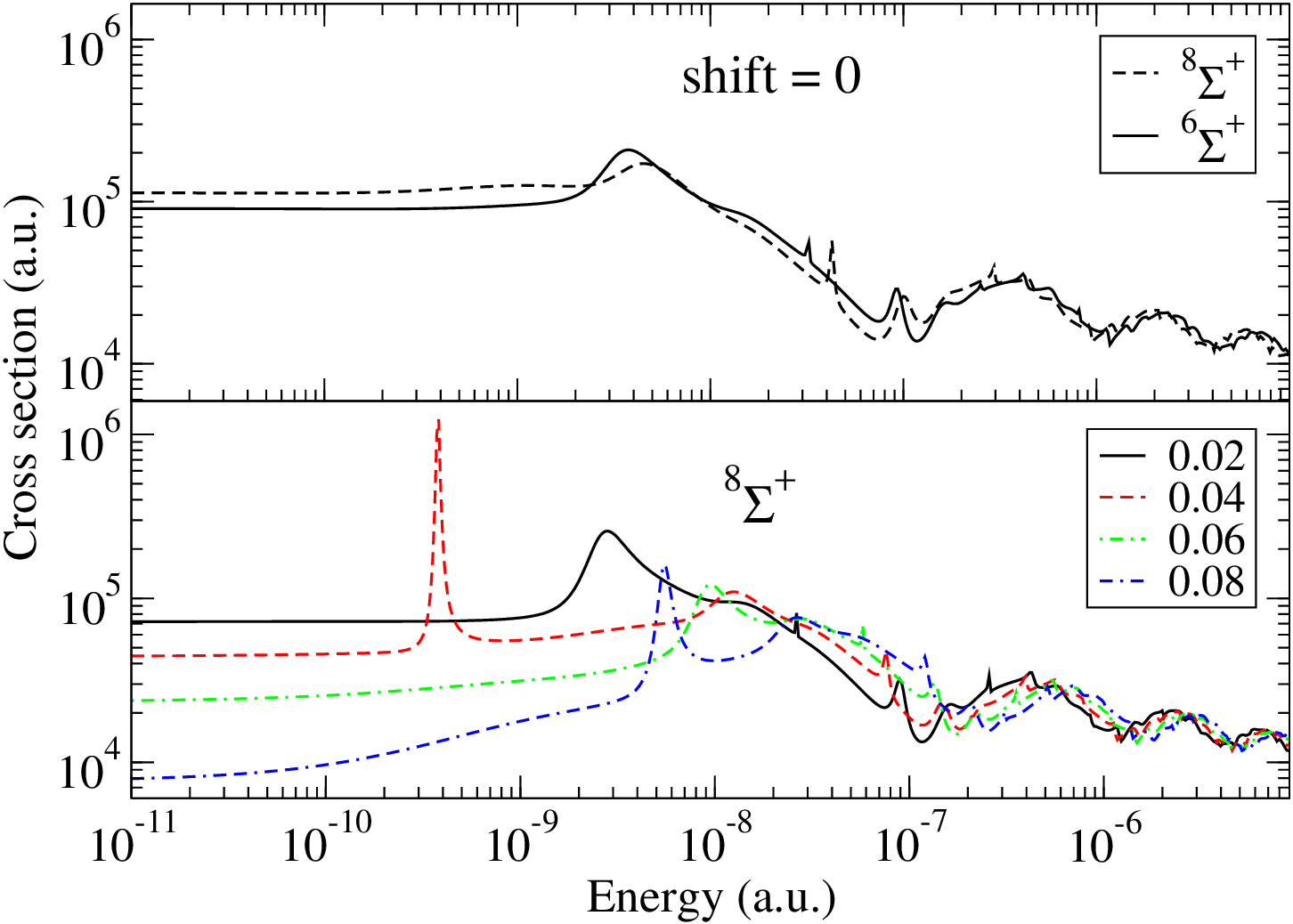}
\caption{\label{shift}
(Color online) Top panel: Elastic cross section for the $^6\Sigma^+$ and $^8\Sigma^+$ 
curves without shift in 
the inner wall $(s=0)$. Bottom panel: Effect of the positive shifts on the elastic cross section 
along  $^8\Sigma^+$  curve.
Potential energy points for which $R < R_e$ 
are shifted according to $R_s = R+s(R-R_e)/(R_t -R_e)$, where $s$ is the shift of the 
zero-energy classical inner turning point, $R_t$, and $R_e$ is the equilibrium 
distance. Several different values 
for $s$ were used.}
\end{figure}
\begin{figure} [ht!]
\includegraphics[width=0.99\columnwidth]{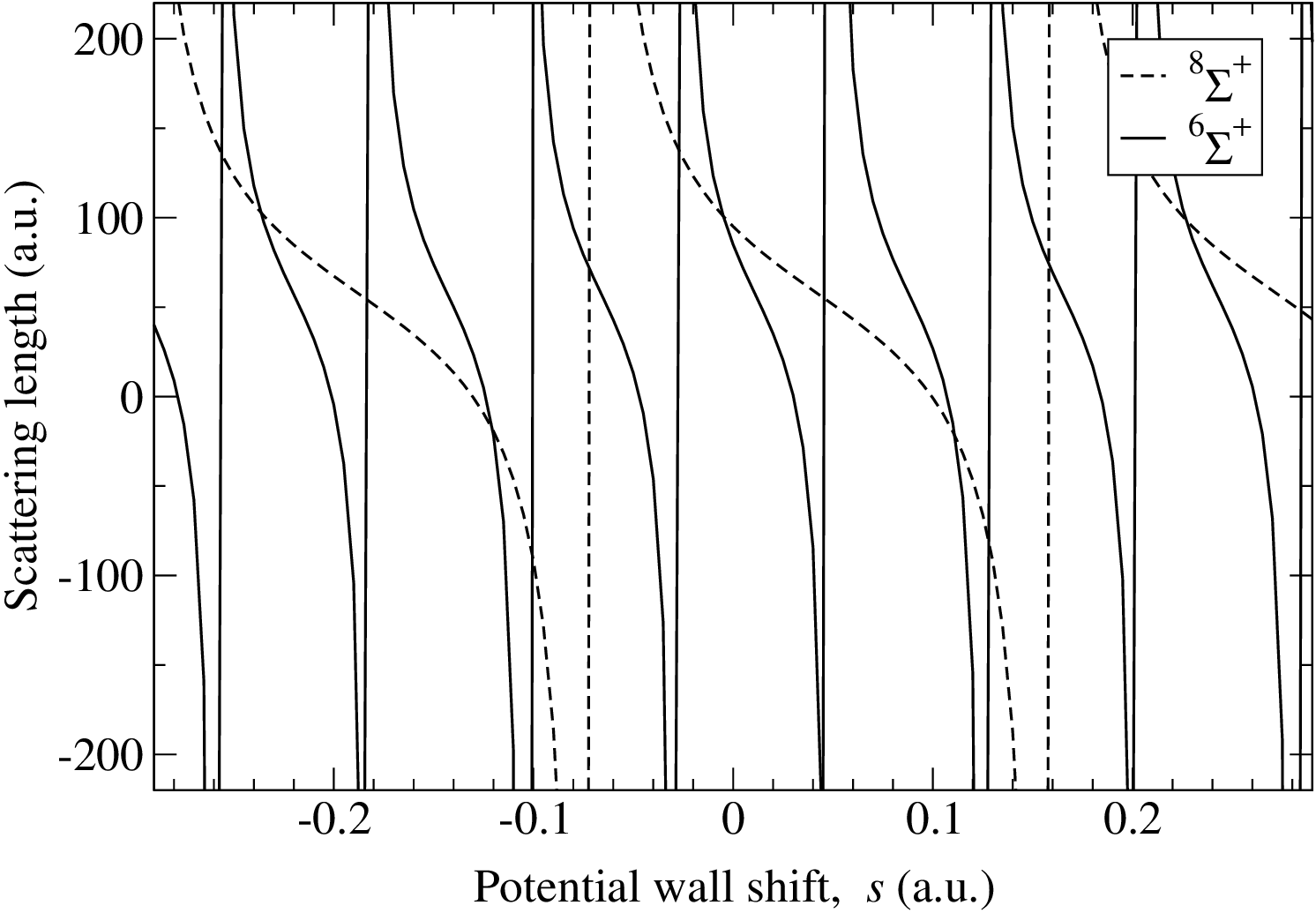}
\caption{\label{bs}
The dependence of the energy of the last bound state in the  $^6\Sigma^+$ and 
$^8\Sigma^+$ potentials on the shift parameter ($s$), see 
Fig.~\ref{shift} for the shift definition. The ``resonances'' occur when an additional bound 
state is pulled in from the continuum as $s$ decreases.}%
\end{figure}

The dearth of spectroscopic and collisional information about the CrRb molecule 
limits what can be extracted from our calculations at ultracold temperatures. Nevertheless, 
we can inform the discussion by modifying the interaction potentials and evidence the changes 
to the phase shifts 
and/or cross sections at very low energies. 
Our vdW coefficient is sufficiently accurate to describe the potentials at large separations.
In the short 
range repulsive wall region, we shift the potential energy curves, according to the prescription 
in our earlier work \cite{pav04}, and the results for the elastic cross sections are shown in 
Fig.~\ref{shift}. 
Several points can be made: while the cross sections are insensitive to the shifts 
for energies above $E\geq 10^{-8}$ a.u., they tend to  decrease (increase)  with 
increasing positive (negative) shifts at very low energies.
However, when the shifts are large enough that an 
additional bound state (see for instance Fig.~\ref{bs}) is pulled in from the continuum, 
the above trend does not hold.
\vspace*{-5mm}
\section{Zeeman cascade}

The scattering between Cr and Rb atoms  in  the presence of 
a homogeneous magnetic field, ${\bm B}$,  is governed by the following Hamiltonian
\begin{align}
\hat{H} =  - \frac{1}{2 \mu R} \frac{\ddpar ^2}{\ddpar R^2} R + 
\frac{{\bm l}^2}{2 \mu R^2} + \hat{V}_{\rm es} + \hat{V}_{\rm dip}  \nonumber \\
 +\;2 \mu_B {\bm s}_{\rm A} \cdot {\bm B} + 2 \mu_B {\bm s}_{\rm B} \cdot {\bm B} 
 + A_{\rm hf} {\bm s}_{\rm B} \cdot {\bm i}_{\rm B},
\label{ham}
\end{align}
\noindent
where the  constants $\mu$, $\mu_B$ and $A_{\rm hf}$ represent  the reduced mass 
of the CrRb molecule, Bohr magneton, and Rb isotropic hyperfine interaction constant 
of the ground state, respectively. 
The first two terms in the Hamiltonian represent the nuclear kinetic energy of the 
molecule, where  ${\bm l}$  is  the rotational angular momentum of the nuclei. 
The explicit form of $\hat{V}_{\rm es}$, the operator of the electrostatic  
interaction, and $\hat{V}_{\rm dip}$, the operator of the magnetic dipolar  
interaction will be given below. The  spin operators  
${\bm s}_{\rm A}$(${\bm i}_{\rm A}$) and ${\bm s}_{\rm B}$(${\bm i}_{\rm B}$) 
represent the electronic (nuclear) spins  of atom A (Cr) and atom B (Rb), 
respectively.
For $^{52}$Cr + $^{87}$Rb system, the electronic and nuclear spins are 
$s_{\rm A} = 3$, $i_{\rm A} = 0$,  
$s_{\rm B} = 1/2$ and $i_{\rm B} = 3/2$
The linear atomic Zeeman terms and the 
 isotropic hyperfine interaction for Rb are also given.
The $R$-independent terms in the Hamiltonian \eqref{ham}, determine the scattering 
channels, which can be related through a unitary transformation 
$\ket{\beta} = \sum_{\alpha} U_{\beta \alpha} \ket{\alpha}$, to the product states 
of the total electronic spin, ${\bm S}  = {\bm s}_{\rm A} + {\bm s}_{\rm B}$, and nuclear spin, 
${\bm I}  = {\bm i}_{\rm A} + {\bm i}_{\rm B}$,  $ \ket{\alpha} = \ket{SM_S IM_I}$.
In this work, we neglect the interaction of the Rb nuclear magnetic moment with $ {\bm B} $.
\par
The total wave function expansion by product states  of  $\tau_{S M_S}$, 
the eigenfunctions of ${\bm S}^2$ and ${\bm S}_z$, and $\tau_{I M_I}$, the eigenfunctions of  
 ${\bm I}^2$ and ${\bm I}_z$, is:
\begin{align}
\Psi\! = \!\sum_{S, M_S}\sum_{I, M_I} G_{S M_S I M_I}({\bm R}) 
 \tau_{S M_S} \tau_{I M_I}.
\label{basis0}
\end{align}
We introduce channel wave functions, $ G_{S M_S I M_I}({\bm R})$, that describe 
the motion of the nuclei. Substitution of the expansion (\ref{basis0}) in the 
Schr\"{o}dinger equation leads to a system of coupled differential equations. 
A further simplification is achieved by introducing a set of 
channel wave functions that depend only on $R$,
\begin{align}
 G_{S M_S I M_I}({\bm R})=\sum_{l, m_l} F_{S M_S I M_I l m_l}(R)Y_{l m_l}(\hat{\bm R}),
\end{align}
where $ \hat{\bm R}$ determines the  direction of ${\bm R}$. 
The  operator $\hat{V}_{\rm es}$ is defined in terms of the 
Born-Oppenheimer interaction potentials $V_S(R)$ for the CrRb molecule,
\begin{align}
\hat{V}_{\rm es}=\sum_{S}\sum_{M_S} V_S(R) |SM_S{\rangle}{\langle}SM_S|.
\end{align}
This decomposition makes  the  operator $\hat{V}_{\rm es}$ diagonal in the basis (\ref{basis0}).
\par
The magnetic dipolar operator \cite{roman1}, in the second rank irreducible tensor 
representation, can be written as
\begin{align}
\hat{V}_{\rm dip} = - \sqrt{\frac{24 \pi}{5}} \frac{\alpha^2}{R^3}
\sum_{q = -2}^{2} (-1)^q Y^{(2)}_{-q} \left [ {\bm s}_{\rm A}
\otimes {\bm s}_{\rm B} \right ]^{(2)}_q,
\label{dipolar}
\end{align}
\noindent
where $\alpha$ is the fine-structure constant.
The matrix of the dipolar interaction (\ref{dipolar}) can be
evaluated analytically as demonstrated, for example, in \cite{roman1,alex}.
It is 
the only term in the Hamiltonian that can couple channel wave functions with different 
rotational numbers, $l$, according to 
\begin{align}
 \Delta l = 0, \pm2; \quad \text{while} \quad  0 \to 0 \quad \text{is forbidden.}
\end{align}
A similar rule holds for the change of the total electronic spin of two atoms
\begin{align}
 \Delta S = 0,\pm1, \pm2.
\end{align}
The dipolar interaction preserves the angular projections, $M_I$, and $ M_S + m_l$  
independently.
\par
The Zeeman interaction term is  $ 2 \mu_0 B M_S $,  
where $M_S$ is the projection of ${\bm S}$ on ${\bm B}$.  
The hyperfine interaction term can be 
directly evaluated in the coupled basis $ \ket{f_{\rm A} f_{\rm B} FM_F}$, 
where ${\bm f}_{\rm A} = {\bm i}_{\rm A} + {\bm s}_{\rm A}$,  
$ {\bm f}_{\rm B} = {\bm i}_{\rm B} + {\bm s}_{\rm B}$, 
and ${\bm F}= {\bm f}_{\rm A} + {\bm f}_{\rm B}$,
and then through the chain of 
transformations 
\begin{align}
 \ket{f_{\rm A}  f_{\rm B} FM_F} \to \ket{SIFM_F} \to \ket{SM_S IM_I},
\end{align}
expressed  in the $ \ket{SM_S IM_I lm_l}$ basis. 
The hyperfine interaction matrix elements are calculated as in \cite{zare,timur}.
\par
\begin{figure}[h!]
\centering
\subfigure{ \label{inelcross}%
\hspace{5pt}\includegraphics[width=0.96\columnwidth]{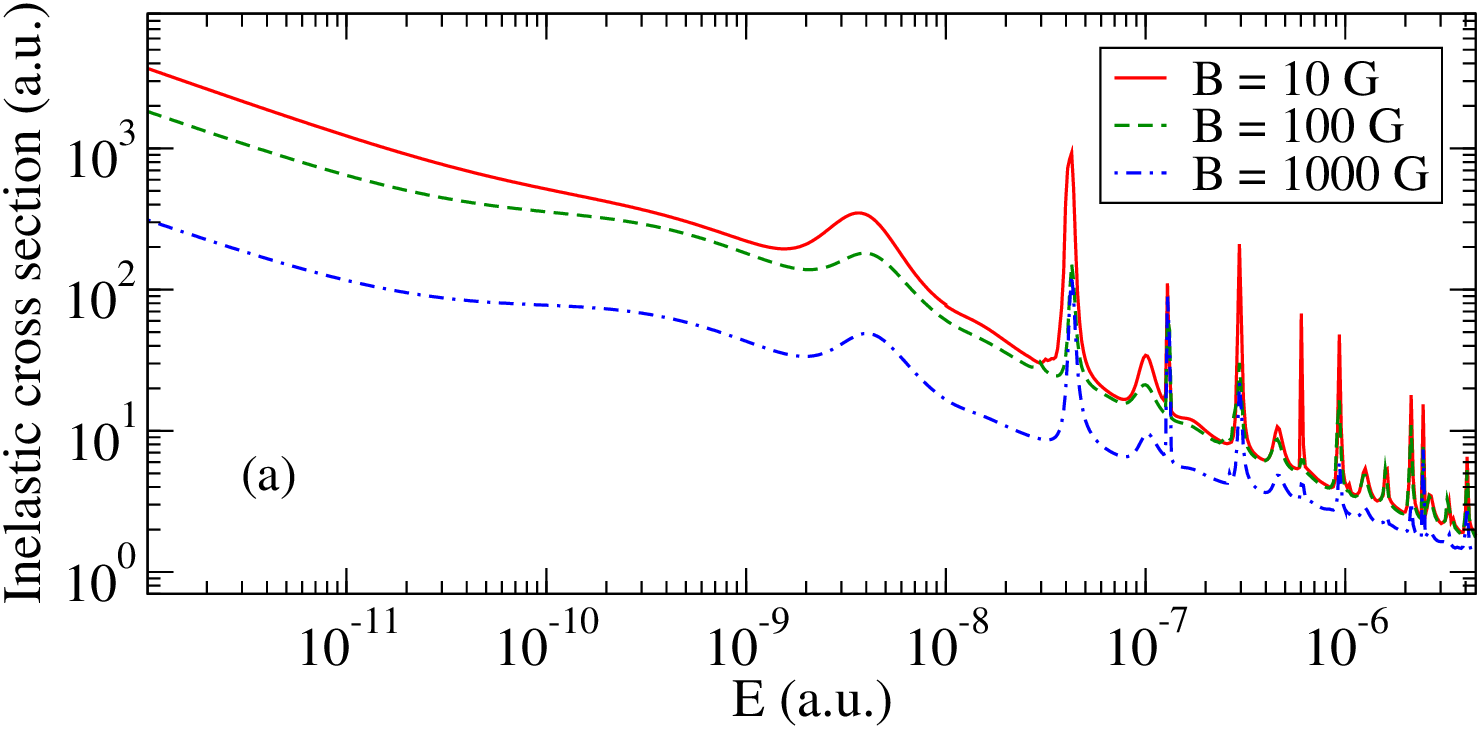}}\\%
\subfigure{ \label{rates}%
\includegraphics[width=0.99\columnwidth]{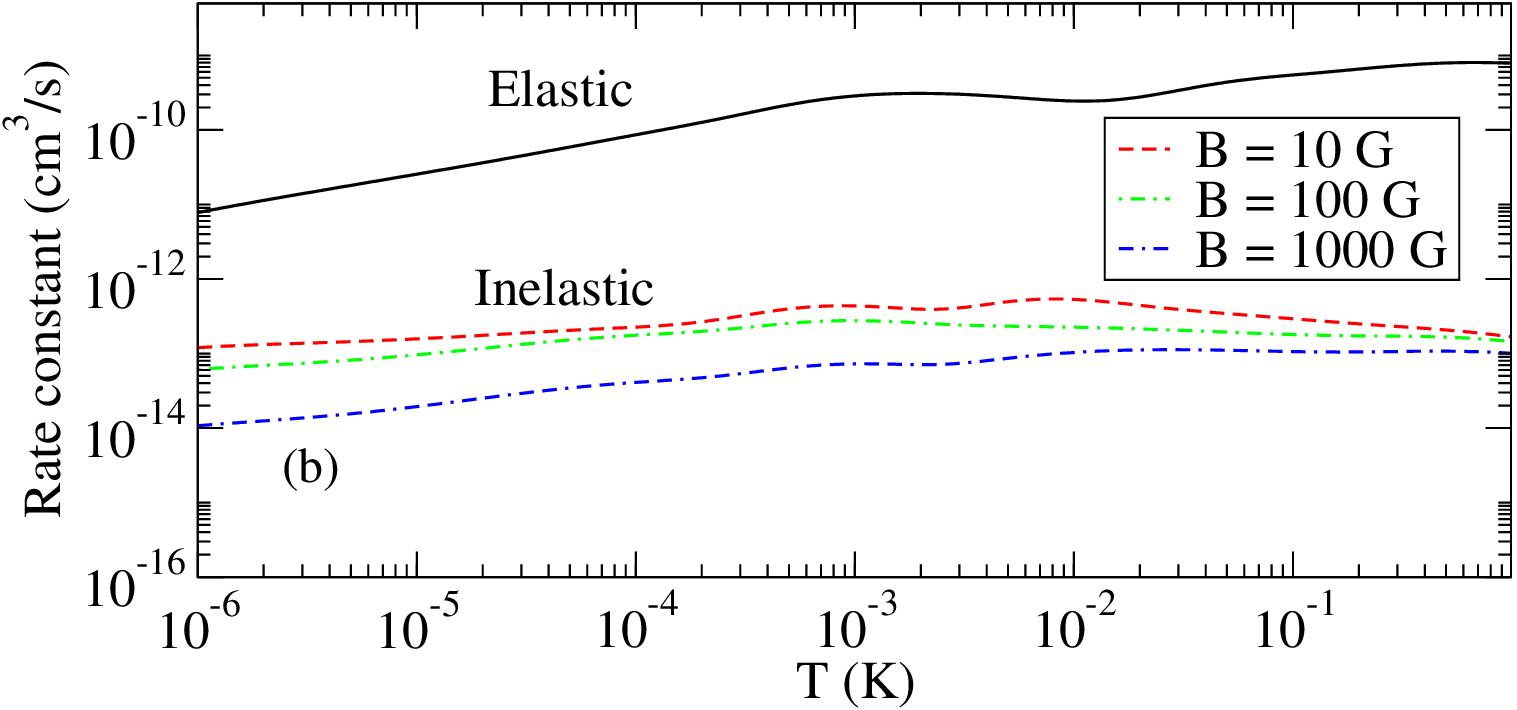}}%
\caption{(Color online) (a) Energy dependence of inelastic cross section for the relaxation of maximally 
stretched $^{52}$Cr, \mbox{$\ket{f{=}3,m_f{=}3}$}, and $^{87}$Rb, \mbox{$\ket{f{=}2,m_f{=}2}$}, 
 atoms in the presence of magnetic field. (b) Rate constants  for elastic and inelastic 
scattering processes presented in  (a).}
\end{figure}

The highest low-field seeking state is chosen as the initial channel  for the Zeeman 
relaxation of maximally stretched Cr and Rb atoms. In this state, the spin 
numbers for $^{52}$Cr are $f=3$ and $m_f = 3$, while for $^{87}$Rb the relevant spin numbers 
are $f=2$ and $m_f$ = 2.
 The results do not change 
much when we exclude states  which represent the decrease of the total spin 
projection, $M_F$, by more than 2 units. 
Figure~\ref{inelcross} contains the  cross sections as a function of  energy at three 
 values of $B$.
\subfiglabelskip=0pt
The cross sections display characteristic dipolar shape resonances, and 
roughly correspond to cross sections for Cr-Cr scattering scaled down 10 times. 
The inelastic cross section contribution from the dipolar interaction 
scales as  \cite{pfau2003}
\begin{align}
 \frac{\sigma_{inel}^{\rm CrRb}}{\sigma_{inel}^{\rm CrCr}} \sim 
 \left(\frac{s_{\rm Rb}}{s_{\rm Cr}} \right)^{3/2} 
 \left( \frac{\mu_{\rm CrRb}}{\mu_{\rm CrCr}} \right)^2.
 \end{align}
 The Zeeman relaxation rates, Fig.~\ref{rates},  
provide a valuable tool in assessing the efficiency  of CrRb evaporative cooling 
in a magnetic or optical trap. Ratios of elastic to inelastic collisions in the $\mu K$ 
 regime of about $100{-}1000$ times are possible.
\par
\begin{figure} [h!]
\includegraphics[width=0.99\columnwidth]{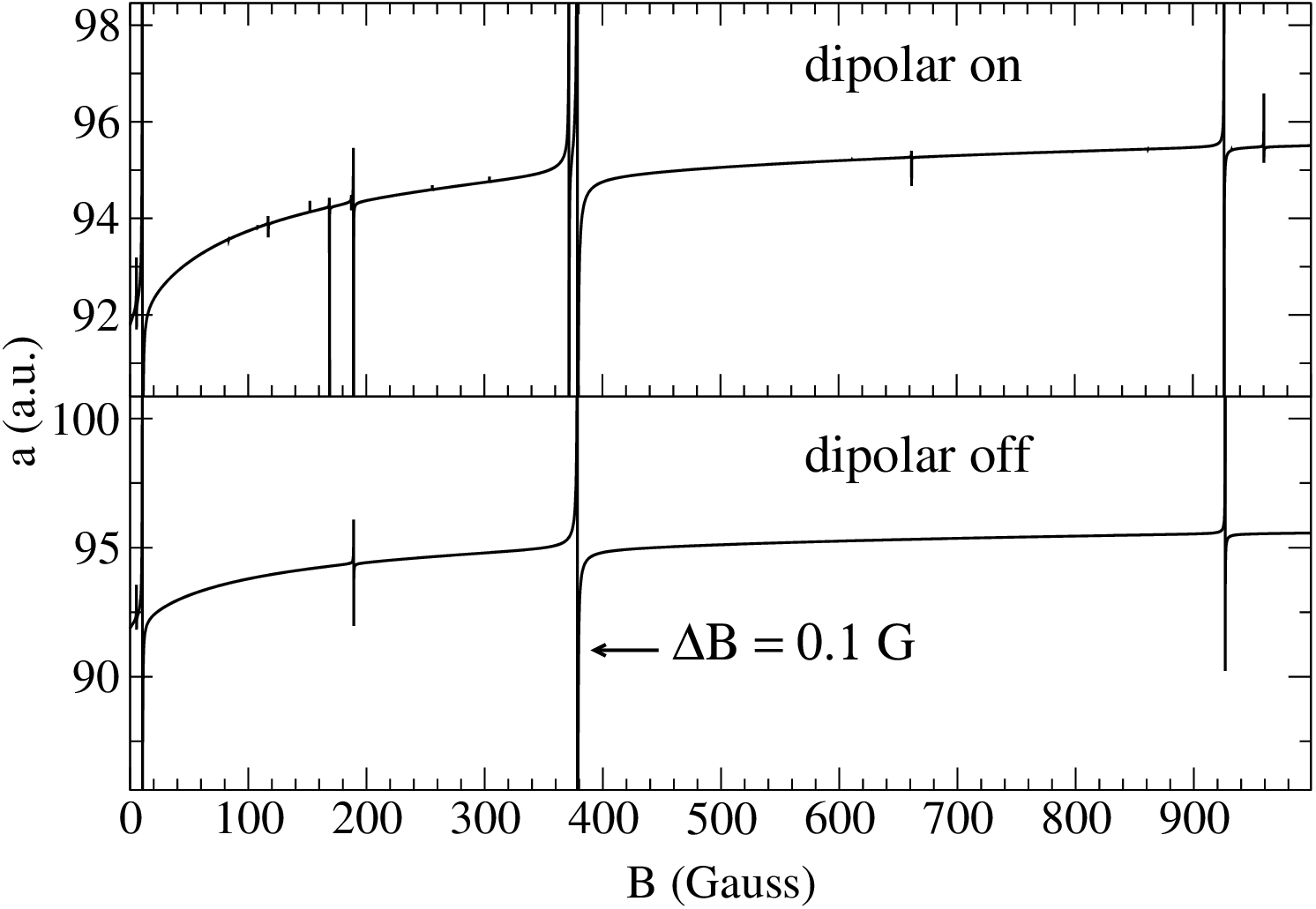}
\caption{\label{hf-dipole}
Competing hyperfine and dipolar interactions in the $l=0$ entrance channel state: 
\mbox{$\ket{f{=}3,m_f{=}{-}3}$} for $^{52}$Cr,  and 
\mbox{$\ket{f{=}1,m_f{=}1}$} for $^{87}$Rb.
 Bottom panel: Resonances caused by Rb hyperfine interaction, 
 Top panel: Effect of dipolar interaction coupling entrance $s$-wave channel to 
 the closed $d$-wave channels.}
\end{figure}

In CrRb, we have an interesting system where strong hyperfine and dipole-dipole interactions 
compete to produce rich spectra. When only the hyperfine interaction in $^{87}\text{Rb}$ 
is considered, the scattering length dependence with the magnetic field, Fig.~\ref{hf-dipole} 
(bottom panel) shows a number of hyperfine-coupled Fano-Feshbach resonances, where the initial 
state is  the lowest high-field seeking state, 
\mbox{$\ket{^{52}\text{Cr}:f{=}3,m_f{=}{-}3;\;^{87}\text{Rb}:f{=}1,m_f{=}1;\; l{=}0,m_l{=}0}$}. 
When the Cr magnetic dipole interaction with the spin of the Rb atom is  ``turned on'', the much 
richer spectrum (top panel in Fig.~\ref{hf-dipole}) 
is obtained, where the additional resonances, due to  the dipolar interaction, have 
($s,d$) wave character.
\par
In summary, we calculate the $^6\Sigma^+$ and  $^8\Sigma^+$ BO potential energy curves of CrRb in the ground 
electronic state, the long-range $C_6$ vdW coefficient, the static  molecular polarizability,  
and the equilibrium permanent electric dipole moment.
We obtain the spin relaxation rate constants for collision of Cr and Rb in a magnetic field and 
investigate the magnetic tunability of the scattering length in the presence of hyperfine and 
dipolar interactions.
With its relatively large electric and magnetic dipole moments, CrRb is a
potentially interesting molecule for collisional studies of dipolar molecules in 
combined electric and magnetic fields. With fields at different angles, it 
may be possible to tune the two dipoles into cooperative or competitive arrangement.

\section*{ACKNOWLEDGMENTS}
This work was supported by a grant from the NSF to ITAMP.
The work of Z.P. and R.C. was partially supported by grant 
PHY-0653449 of the NSF. 

\end{document}